\documentstyle[11pt,paspconf,psfig]{article}
\begin{document}
\title{Collisional Ring Galaxies in Small Groups}
\author{C. Horellou}
\affil{Onsala Space Observatory, Chalmers University of Technology,
S-439 92 Onsala, Sweden}
\def\HI{H{\sc i}\,}
\begin{abstract}
The probability of plunging orbits is enhanced in groups
of galaxies and indeed, observations show that ring galaxies,
which are believed to form when a galaxy passes through
the center of a larger rotating disk, are often found in 
small groups. Numerical simulations combined with a knowledge of the 
large-scale \HI distribution 
provide strong constraints on the 
dynamical history of these systems 
and on the identity of the intruder. 
Here we present a numerical model of the Cartwheel
which supports the suggestion that the most distant companion is the
intruder. We also present 
high-resolution H{\sc i}\, observations of the more irregular system Arp 119 
that reveal a possible connection
to the most distant companion. 

\end{abstract}

\keywords{}

\section{Introduction}   

The evolution of galaxies in small groups is 
governed by gravitational interaction. 
Even if the lifetime of a dynamical system
is 5--10 times larger than the crossing time, small groups of
galaxies should have merged into one single galaxy in much less than 
a Hubble time. It has been suggested that small groups  
are constantly replenished 
through infall of galaxies in  looser groups 
(Diaferio, Geller \& Ramella 1994). Another scenario
supported by  
numerical simulations is that dark matter halos 
may delay the merging process of small groups, 
in particular if the halo  
envelopes the whole group
(Barnes 1985; Athanassoula et al. 1997). 
 
The presence of a ring galaxy in a group may help provide
constraints to a number of parameters, such as the 
dynamical timescales and the dark matter distribution. 
Rings form after the passage of a smaller galaxy through the center 
of a disk galaxy. 
They are expanding ring waves made more visible through the blue light
of the young stars born from the gas that has been compressed in the ring
wave. 
From the size and the expansion velocity of the ring (as measured for 
instance by high-resolution emission line observations), 
it is possible
to estimate the age of the ring. The relative spacing
of the rings gives information on the dark matter distribution in
that galaxy. The large-scale atomic gas distribution traces  
the history  of the interaction and, combined with
numerical N-body simulations, provides some insight into the evolution
of those groups. One of the best studied northern ring systems is 
VIIZw466 for which 
\HI observations have 
provided information on the kinematics
of the gas in the ring and its nearby companions 
and have revealed the presence of a plume extending from one
of the companions towards the ring 
(Appleton, Charmandaris
\& Struck 1996). The effect of an asymmetric compression in the ring 
was seen in the mid-infrared (using $ISO$) and the radiocontinuum emission  
(Appleton, Charmandaris, Horellou et al. 1999). 
Here we discuss two more ring galaxies located in small  
groups: the Cartwheel and Arp 119; we present a model of the Cartwheel
that reproduces most of the observed features, and the first high-resolution
\HI observations of Arp 119.

\section{A Model for the Cartwheel}    
\begin{figure}[]
\centerline{\psfig{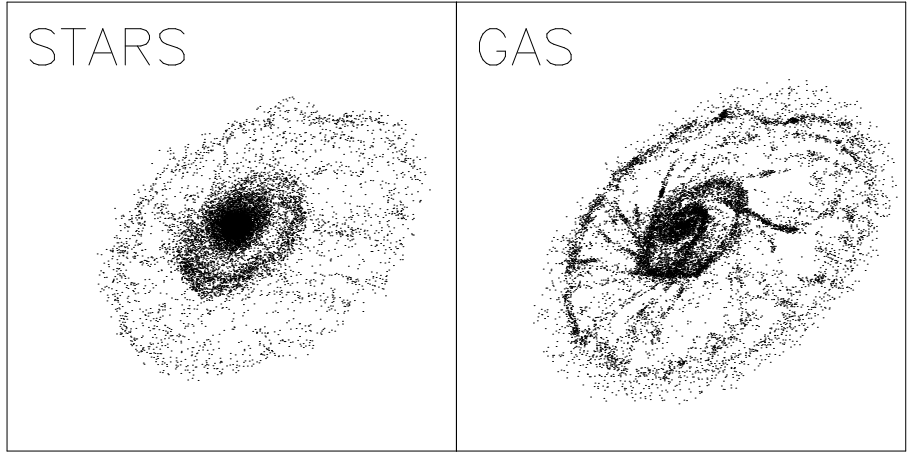}}
\caption{Numerical simulation of the Cartwheel. The intruder 
(not visible on this figure) 
lies on the plane of the sky close to G3, the most distant
companion of the Cartwheel }
\end{figure}

\begin{figure}[t]
\centerline{\psfig{file=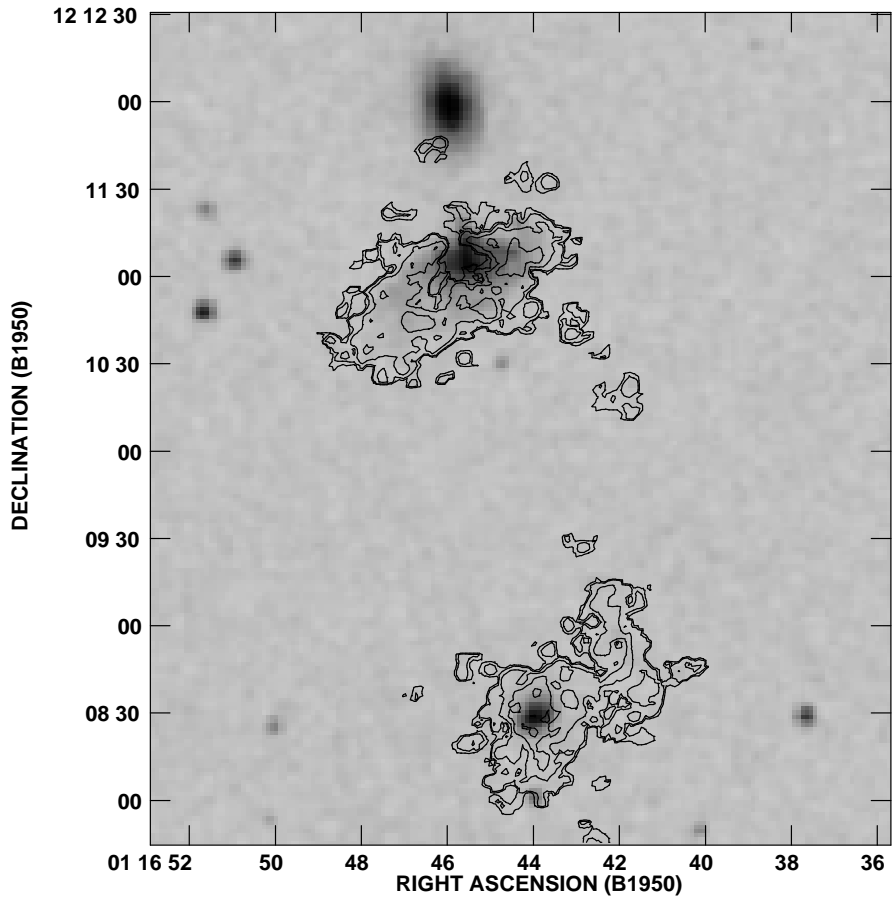,width=12cm}}
\caption{Atomic gas distribution in Arp 119 (contours: 
VLA B-array total \HI intensity map; grey-scale: optical image from the
Digitized Sky Survey). The angular resolution of the \HI map is 10$''$.} \label{fig-1}
\end{figure}

The results of our simulation assuming the Cartwheel's most distant 
companion (called G3) as the intruder are shown in Fig. 1.    
We choose G3 as the ``bullet" rather than one of the two nearer  
companions because of the existence of an extended \HI plume 
reaching from the Cartwheel towards that galaxy 
(Higdon 1996). 
Our model is based on the nearly central collision of a rigid companion
galaxy with a self-gravitating disk containing both stars and gas 
(see Horellou \& Combes 2000a for more details).  
The halo of our computer-made Cartwheel is about three time as massive
as the disk+ bulge, and the mass of the companion is half that of the
target. 
The model reproduces the main features of the Cartwheel 
and the observed position and radial velocity difference between
the Cartwheel and G3.   
It is interesting than a good fit can also be obtained by assuming 
one of the nearer companions as the intruder 
on a trajectory that is perpendicular to that of the Cartwheel's disk 
(Bosma et al., this volume).  However, neither Bosma et al.'s model nor
ours involving a rigid intruder is able to reproduce the $\approx$ 100 kpc 
long \HI tail towards G3.  
In a more elaborate calculation in which both the companion and target are
deformable, we found that a gas-rich companion plunging through the center
of the target on a prograde orbit can extrude a gaseous plume reminiscent of 
that in the Cartwheel 
(Horellou \& Combes 2000b).  

\section{Arp 119}  

Arp 119 is a more distorted system with two nearby companions, an 
elliptical and an irregular. 
In addition to the material directly associated with the galaxies, 
\HI emission is detected south of the ring-like one and north
of the southern irregular companion Mrk 983 
(see Fig. 2; Horellou, Charmandaris \& Combes, in prep). 
The two features may be connected, 
forming a bridge between the two galaxies. The \HI in Arp 119 is distributed
in a broad rotating ring with several condensations, some of which coincide 
with the knots terminating the ``spokes" that are visible on the optical
picture. Those features may be due to the head-on collision between 
Mrk 983 and Arp 119.  The nucleus of Arp 119  presents characteristics
of a LINER. It is not clear whether the head-on collision of two gas-rich
galaxies can trigger an active nucleus, 
although the fraction of ring galaxies with a Seyfert nucleus is rather 
high (four out of about 30). 

\section{Conclusions}

As originally shown by Few \& Madore (1986), ring galaxies which  
present signs of having undergone a collision have, on the average, 
a higher number of neighbors than galaxies in a control sample.    
When two companions are present, \HI observations help
reveal the identity of the intruder.   X-ray maps of groups containing
a ring would give further constraints on the degree of relaxation of the
group. So far, 
the Cartwheel is the only ring galaxy in which X-ray emission has been detected 
(ROSAT observations by Wolter, Trinchieri \& Iovino 1999) and the emission
is concentrated in the southern ring quadrant where most of the star formation
occurs.  The VIIZw466 system -- which contains an asymmetric ring produced in 
an off-center collision, an edge-on spiral and a massive ellipical with 
distorted isophotes that may have accreted several companions --
would be an ideal target for observations with 
Chandra. 

\acknowledgments
The author enjoyed discussions with 
Vassilis Charmandaris, Fran\c{c}oise Combes and Phil Appleton 
on ring galaxy projects. Special thanks to John and Steven Black
for their support and their interest in telescopes and galaxies.


\begin{references}
\reference	Appleton, P.N., Charmandaris, V. \& Struck, C., 
1996, \apj, 468, 532 
\reference	Appleton, P.N., Charmandaris, V., Horellou, C., 
Mirabel, I.F., Ghigo, F., Higdon, J.L., Lord, S., 1999, \apj, (in press) (astro-ph/9907122)
\reference 	Athanassoula, E., Makino, J., Bosma, A., 1997, \mnras, 286, 825
\reference 	Barnes, J.E., 1985, \mnras, 215, 517
\reference	Diaferio, A., Geller, M.J., Ramella, M., 1994, \aj, 107, 868
\reference	Few, M.A., Madore, B.F., 1986, \mnras, 222, 673 
\reference      Higdon, J.L., 1996, \apj, 467, 241
\reference	Horellou, C., Combes, F., 2000a, Astrophysics \& Space
     Science, in press (Proceedings of the Euroconference on 
The Evolution of Galaxies on Cosmological Timescales held in
Tenerife, Dec. 1998. Eds Beckman, J.E., Mahoney, T.) 
\reference	Horellou, C., Combes, F., 2000b, ASP Conference Series,
15th IAP Meeting held in Paris, France, July 9-13, 1999, in press, 
Eds.: Combes, F.,
Mamon, G.A., and Charmandaris, V., Galaxy Dynamics: from the Early Universe
to the Present
\reference	Horellou, C., Charmandaris, V., Combes, F.,  to be submitted 
to A\&A 
\reference	Wolter, A., Trinchieri, G., Iovino, A., 1999, A\&A, 342, 41
\end{references}
\end{document}